\documentclass[]{article}

\usepackage{ragged2e}
\usepackage{graphicx} 
\usepackage{wrapfig}
\usepackage{amssymb}
\usepackage{amsmath}
\usepackage{dirtytalk}
\usepackage{xcolor}
\usepackage{bm}
\usepackage{amsthm}
\usepackage{float}
\theoremstyle{plain}
\usepackage[margin=1in]{geometry}
\newtheorem{assumption}{Assumption}
\usepackage[authoryear]{natbib}
\usepackage{algpseudocode}
\usepackage{comment}
\newcommand{\myvar}[1]{{#1}}
\newcommand{\myset}[1]{\mathcal{#1}}
\usepackage{subcaption}

\begin{document}

\title{Supervisory Control of Hybrid Power Plants Using Online Feedback Optimization: Designs and Validations with a Hybrid Co-Simulation Engine}

\author{Sayak Mukherjee$^{1\dagger}$, Himanshu Sharma$^{1\dagger}$, Wenceslao Shaw Cortez$^{1,a\dagger}$, Genevieve Starke$^{2}$, \\ Michael Sinner$^{2}$ , Brooke J. Stanislawski$^{2}$, Zachary Tully$^{2}$, Paul Fleming$^{2}$, Sonja Glavaski$^{1}$
\thanks{ Affiliations:$^1$ Pacific Northwest National Laboratory, Richland, WA 99352, USA, $^2$ National Renewable Energy Laboratory, Golden, CO, USA. Corresponding email: \{sonja.glavaski\}@pnnl.gov, $^\dagger$contributed equally, $^a$W.S. Cortez was with the Pacific Northwest National Laboratory while contributing to this work. }}%

\date{}

\maketitle

\vspace{-0.6 cm}

\begin{abstract}
This research investigates designing a \textit{supervisory feedback} controller for a hybrid power plant that coordinates the wind, solar, and battery energy storage plants to meet the desired power demands. We have explored an online feedback control design that does not require detailed knowledge about the models, known as \textit{feedback optimization}. The control inputs are updated using the gradient information of the cost and the outputs with respect to the input control commands. This enables us to adjust the active power references of wind, solar, and storage plants to meet the power generation requirements set by grid operators. The methodology also ensures robust control performance in the presence of uncertainties in the weather. In this paper, we focus on describing the supervisory feedback optimization formulation and control-oriented modeling for individual renewable and storage components of the hybrid power plant. The proposed supervisory control has been integrated with the hybrid plant co-simulation engine, Hercules, demonstrating its effectiveness in more realistic simulation scenarios. 

\end{abstract}



\section{Introduction}
\label{sec:intro}

Hybrid power plants \citep{walton2019nextera, paska2009hybrid} that combine renewable generation from wind farms and solar plants with energy storage systems are envisioned to be one of the major contributors to energy sector reliability and security, both in the United States and globally. With declining costs of wind, solar, and energy storage technologies, hybrid systems are poised to reduce the expenses associated with expanding renewable deployment \citep{haegel2017terawatt, dykes2019results}. Many of the hybrid plant designs include utilizing wind-based designs where co-located wind and other resources make the resource mix for such plants.  The mixed resources in the wind-based hybrid plant help decrease the uncertainty caused by only a wind-based renewable plant, where in most of the scenarios, energy storage units are utilized. The stable and consistent energy delivery of the renewable plants is constrained by wind profiles \citep{roy2010optimum} and solar variability \citep{rosenkranz2016analyzing}. However, hybrid wind-solar-storage plants can capture complementarity between wind and solar resource profiles and provide a buffer for the variable resources, which helps to improve reliability and smooth the plant power output compared to stand-alone wind or solar generation \citep{pan2009dynamic, tina2006hybrid}. There are a plethora of works that look into designing the optimal plant, starting from location selection to sizing and dispatchability considerations \citep{kabalci2013design, stanley2022optimizing}. Along with sizing-based optimization, the hybrid plant operator is also tasked with maintaining reliable generation supply to meet the required load demand in both long-term and near-term horizons. This necessitates developing supervisory controllers that operate on a relatively faster timescale, from seconds to minutes, such that the resources could coordinate between themselves to support the required power demand, which subsequently may be beneficial in frequency support and other grid-supporting services.

Conventionally, plant operators solve an offline optimization problem to schedule dispatches \citep{zurita2021multi, ming2018optimal} of individual power generation or energy storage units. The dispatch process relies heavily on accurate power generation predictions that are based on weather forecasts and is extremely sensitive to modeling errors and external disturbances. Moreover, a substantial amount of data is required to develop power production models dependent on weather prediction. 
Literature in control methods remains
mostly focused on model-based designs, use of heuristics or scheduled dispatches solved offline. In studies \citep{li2013battery, alam2014novel}, power smoothing control has been suggested for energy storage systems to reduce power fluctuations from solar and wind plants. \citep{sun2019optimal} shows a coordinated power control for pumped hydropower units. Recent research has also explored coordination among resources for coordinated control in various hybrid configurations such as wind-storage \citep{bullich2017active}, solar-storage \citep{miao2015coordinated}, and wind-solar-storage systems \citep{pombo2019novel}. \citep{shen2020extremum} presents an extremum seeking based supervisory control for a hybrid DC system. \citep{paduani2022optimal} shows MPC-based optimal supervisory design for a PV-based hybrid power plants utilizing the detailed sub-component models. A switched system strategy based control of hybrid plants using the detailed dynamic model has been discussed in \citep{noghreian2020power}. Heuristic algorithms such as adaptive neuro-fuzzy based  MPPT techniques are also utilized in literature for control of hybrid plants \citep{kechida2024smart}. Therefore, most of the hybrid plant controllers use model-based optimal scheduling or heuristic approaches that do not have guarantee to adapt quickly to disturbances, and may require re-computing the dispatch schedule with sufficient changes in the resource availability. This necessitates the need for embedding the optimization in the feedback loop that can continuously update the optimal supervisory set-points as the system conditions evolve.

This paper introduces a \textit{supervisory} control of the hybrid power plants that, without knowing accurate power production models, calculates in real time the power set points for wind, solar, and battery plants, enabling the hybrid plant to meet the required power demands. The supervisory controller  utilizes feedback from real-time power generation of individual renewable plants to update the control inputs. The supervisory controller generated power set points are subsequently provided to low-level control laws. The low-level controllers are used in the wind, solar, and battery plants, respectively. For example, the low-level control for the wind module would take the desired power set point and control the wind farm (e.g., pitch control, torque control) to generate power that matches the desired power set point. Similar low-level controllers are assumed to be implemented for the solar farm and battery module.    

\begin{figure}[!t]
    \centering
    \includegraphics[width = 0.6\linewidth]{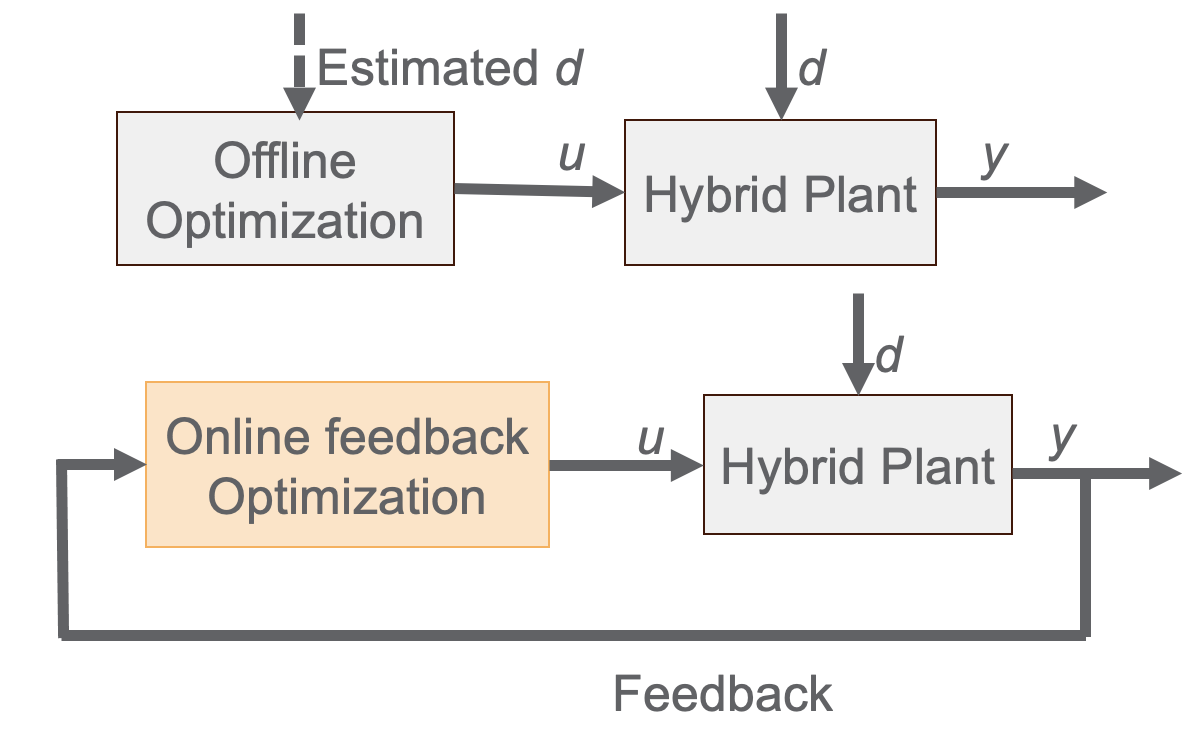}
    \caption{Implementations of offline optimization (top) and online feedback optimization (bottom) control for hybrid plant, where \textbf{d} is demand, \textbf{u} are the control inputs and \textbf{y} are the plant measurements. }
    \label{fig:schema}
\end{figure}
We use a recently researched feedback control strategy, referred to as feedback optimization \citep{Allibhoy2023, Ortmann2023, Colot2023, Picallo2022}, which is robust to uncertainties in the weather. Feedback optimization is a method that solves an optimization problem with the system in-the-loop. The cost function in this optimization is used to match the power demands while respecting individual hybrid power plant component constraints, e.g., max power limits and battery charging limits. Figure \ref{fig:schema} shows the differences in implementations of this approach from the offline optimization-based dispatch solutions where the main benefit is obtained by utilizing real-time feedback interconnection.  

In the proposed approach, the feedback optimization control will operate at a slower (minutes/seconds) timescale than the low-level controllers and subsystem (wind, solar, battery) dynamics. This allows for modeling the subsystems as input-output mappings without the requirement to include the complex low-level closed-loop dynamics. Additionally, assuming changes in the weather are relatively slow, the feedback optimization is made robust to slowly varying disturbances of the hybrid power plant component input-output mappings. This effectively enables a supervisory control that provides power demands while accounting for uncertainties in the weather forecasting. For the implementation aspect of the supervisory control, we have developed different components of the Wind Hybrid Open Controller (WHOC) for hybrid power plants~\citep{WHOC_2024}. 

In this paper, we present a supervisory control in the form of a feedback optimization control law and set up the control-oriented model to define the controller for implementation. Section \ref{sec:background_hybrid} describes the background on the hybrid power plants, and feedback optimization control design. Section \ref{sec:problem} formulates the supervisory control for the hybrid plants that meets the required power demands. Section \ref{sec:solution} describes the proposed solution with details of the control-oriented modeling of individual plants required to implement the online feedback optimization controller, along with numerical examples with realistic resource profiles with simplified steady-state models. Subsequently, the details on the integration of the supervisory controller with the hybrid co-simulation engine are described in Section \ref{sec:high_fidelity} along with numerical test results. Concluding remarks are provided in Section \ref{sec:conclusion}. 

\section{Background}
\label{sec:background_hybrid}
\subsection{Hybrid Power Plants}
Hybrid power plants are combined assets from multiple energy resources coupled together to support grid demands and provide necessary services as required. The hybrid resources are intended to provide predictable and controllable electricity generation. Wind-based hybrid power plants utilize large-scale wind generation as the primary resource supported by solar PV units and storage units in general. The primary form of hybrid power plants intends to produce electricity to support grid operations, whereas there are other forms of hybrid plants that can also produce fuel such as hydrogen. This work considers the first type of hybrid energy resource where wind, solar, and energy storage units are considered to be co-located, and we are tasked with developing the coordinating control architecture such that they are operationally coupled. 

Conventional energy generation from synchronous generators provides reliable and dispatchable supply, which renewable-resource-based plants struggle with. For improving the steady supply of renewable resources, the demand side or the supply side can be controlled. Hybrid plants focus on making the supply side more consistent to support the demands. Hybrid resources, when fully coordinated, can utilize the complementary nature of resources to make the total power output from the hybrid plant more controllable. When integrated with energy storage units, the available capacity is usable by minimizing uncertainties associated with wind and solar resources. The problem we consider in this paper is to support the power demand signal received by the plant operator in the near-term time horizons with the help of real-time feedback of the generators and power demand signal.

\subsection{Mathematical Preliminaries to Feedback Optimization}\label{ssec:fo}
Feedback optimization is a type of control law that uses a system's input-output mapping to converge to the solution of an optimization problem \citep{Allibhoy2023, Ortmann2023, Colot2023, Picallo2022}. In this sense, the measurements from the system (represented by input-output mapping) are fed back into the control law to converge to the optimal solution, hence the name ``feedback optimization." 

First, consider the system represented as an input-output mapping defined by $h: \mathbb{R}^{n_u} \times \mathbb{R}^{n_d} \to \mathbb{R}^{n_y}$:
\begin{equation}\label{eq:input-output mapping}
    \myvar{y} = h(\myvar{u}, \myvar{d})
\end{equation}
where $\myvar{y} \in \mathbb{R}^{n_y}$ is the system output, $\myvar{u} \in \mathbb{R}^{n_u}$ is the system input (that will be calculated by the controller), and $\myvar{d} \in \mathbb{R}^{n_d}$ is an uncertain disturbance acting on the system.

Next, to determine control action, we define  the following optimization problem with respect to the system's input-output map, Eq. \eqref{eq:input-output mapping}:

\begin{subequations}\label{eq:fo main problem}
\begin{align}
 \myset{P}: \hspace{0.4cm} &  \underset{\myvar{u}} {\text{min}} \hspace{.3cm} \ell \left(h(\myvar{u}, \myvar{d}), \myvar{u}\right)   &&\\
 & \hspace{0.1cm} \text{s.t. }  \nonumber&&\\
&  \hspace{0.9cm} g\left(h(\myvar{u}, \myvar{d}), \myvar{u}\right) \leq 0 &&\\
&\hspace{0.9cm}  c(h(\myvar{u}, \myvar{d}), \myvar{u}) = 0
\end{align}
\end{subequations}
where $\ell: \mathbb{R}^{n_y} \times \mathbb{R}^{n_u} \to \mathbb{R}$ is the cost function, $g: \mathbb{R}^{n_y} \times \mathbb{R}^{n_u} \to \mathbb{R}^{n_g}$ is the inequality constraint function, and $c: \mathbb{R}^{n_y} \times \mathbb{R}^{n_u} \to \mathbb{R}^{n_g}$ is the equality constraint function. It is important to note that if $\myvar{d}$ is known exactly, then $\myset{P}$ can be solved directly using existing nonlinear programming solvers. However, in practice, the disturbance on a system is often unknown or uncertain. In the case of a hybrid plant, for example, uncertainties arise in wind predictions, wake disturbances, and/or solar occlusions, which can yield suboptimal solutions to Eq.~\eqref{eq:fo main problem}.

In order to ensure $\myset{P}$ is well-defined, we make the following assumption:
\begin{assumption}\label{asm:FO assumptions}
 $\myset{P}$ is feasible and let $\myvar{u}^* \in \mathbb{R}^{n_u}$ be a local minimizer and isolated KKT point of $\myset{P}$ for a given $\myvar{d}$. Assume the following conditions hold: 
    \begin{enumerate}
        \item The mappings $\myvar{u} \mapsto \ell(h(\myvar{u}, \myvar{d}),\myvar{u})$ and $\myvar{u} \mapsto h(\myvar{u}, \myvar{d})$ are twice-continuously differentiable over an open neighborhood of $\myvar{u}^*$ and their Hessian matrices are positive definite at $\myvar{u}^*$.
        \item The mappings $\myvar{u} \mapsto g(h(\myvar{u}, \myvar{d}),\myvar{u})$ and $\myvar{u} \mapsto c(h(\myvar{u}, \myvar{d}),\myvar{u})$ are continuously differentiable.
        \item The gradient of $h$ with respect to $\myvar{u}$ is not dependent on $\myvar{d}$: $\nabla_u h(\myvar{u}) \equiv \nabla_u h(\myvar{u}, \myvar{d})$.
        \item The strict complementarity condition and the linear independence constraint qualification hold at $\myvar{u}^*$ \citep{Nocedal2006}.
    \end{enumerate}
\end{assumption}

In order to solve Eq. \eqref{eq:fo main problem}, feedback optimization applies the following control law to Eq. \eqref{eq:input-output mapping}:
\begin{equation}\label{eq:FO control}
    \dot{\myvar{u}}(t) = \eta F(\myvar{y}(t), \myvar{u}(t)), \ \myvar{u}(0) = \myvar{u}_0
\end{equation}
\begin{subequations}\label{eq:fo F optimization}
\begin{align}
 F(\myvar{y}, \myvar{u}) = \hspace{0.1cm}&  \underset{\myvar{\theta}} {\text{argmin}} \hspace{.1cm} \|\myvar{\theta} + (\nabla_u h(\myvar{u})^T \nabla_y \ell(\myvar{y}, \myvar{u})^T + \nabla_u \ell(\myvar{y}, \myvar{u})) \|_2^2   &&\\
 & \hspace{-1cm} \text{s.t. }  \nonumber&&\\
&  \hspace{-1cm} (\nabla_u h(\myvar{u})^T \nabla_y g (\myvar{y}, \myvar{u})^T  + \nabla_u g(\myvar{y}, \myvar{u})^T) \myvar{\theta} \leq -\beta g(\myvar{y}, \myvar{u}) &&\\
&\hspace{-1cm} ( \nabla_u h(\myvar{u})^T \nabla_y c(\myvar{y}, \myvar{u})^T + \nabla_u c(\myvar{y}, \myvar{u}) ^T) \myvar{\theta} = -\beta c(\myvar{y}, \myvar{u}) &&\\
&\hspace{-1cm}  w(\myvar{\theta}) \leq 0
\end{align}
\end{subequations}
where $\myvar{u}_0 \in \mathbb{R}^{n_u}$ is the initial control value and $\eta \in \mathbb{R}_{>0}$, $\beta \in \mathbb{R}_{>0}$ are gain terms, and $w:\mathbb{R}^{n_\theta} \to \mathbb{R}^{n_w}$ are rate constraints to impose restrictions on  $\dot{\myvar{u}}$. The main idea behind this control law is to exploit the gradient of $h$, which is independent of $\myvar{d}$, to converge to the solution of $\myset{P}$ instead of attempting to solve $\myset{P}$ directly.  

If Assumption \ref{asm:FO assumptions} holds, then the results from \citep{Allibhoy2023, Colot2023} ensure that the control defined by Eqs. \eqref{eq:FO control} and \eqref{eq:fo main problem} converges to the minimizing argument of $\mathcal{P}$, and if the constraints are satisfied at the initial time, i.e., $g(\myvar{y}(0), \myvar{u}(0)) \leq 0$, $c(\myvar{y}(0), \myvar{u}(0)) = 0$, then the constraints will be satisfied at all times. Furthermore, the results show that if at an initial time the constraints are \emph{not} satisfied, then the control will direct $\myvar{y}$ and $\myvar{u}$ towards the safe region.

\section{Supervisory Control Problem Formulation}\label{sec:problem}

\begin{figure}[!t]
    \centering
    \includegraphics[width = \linewidth]{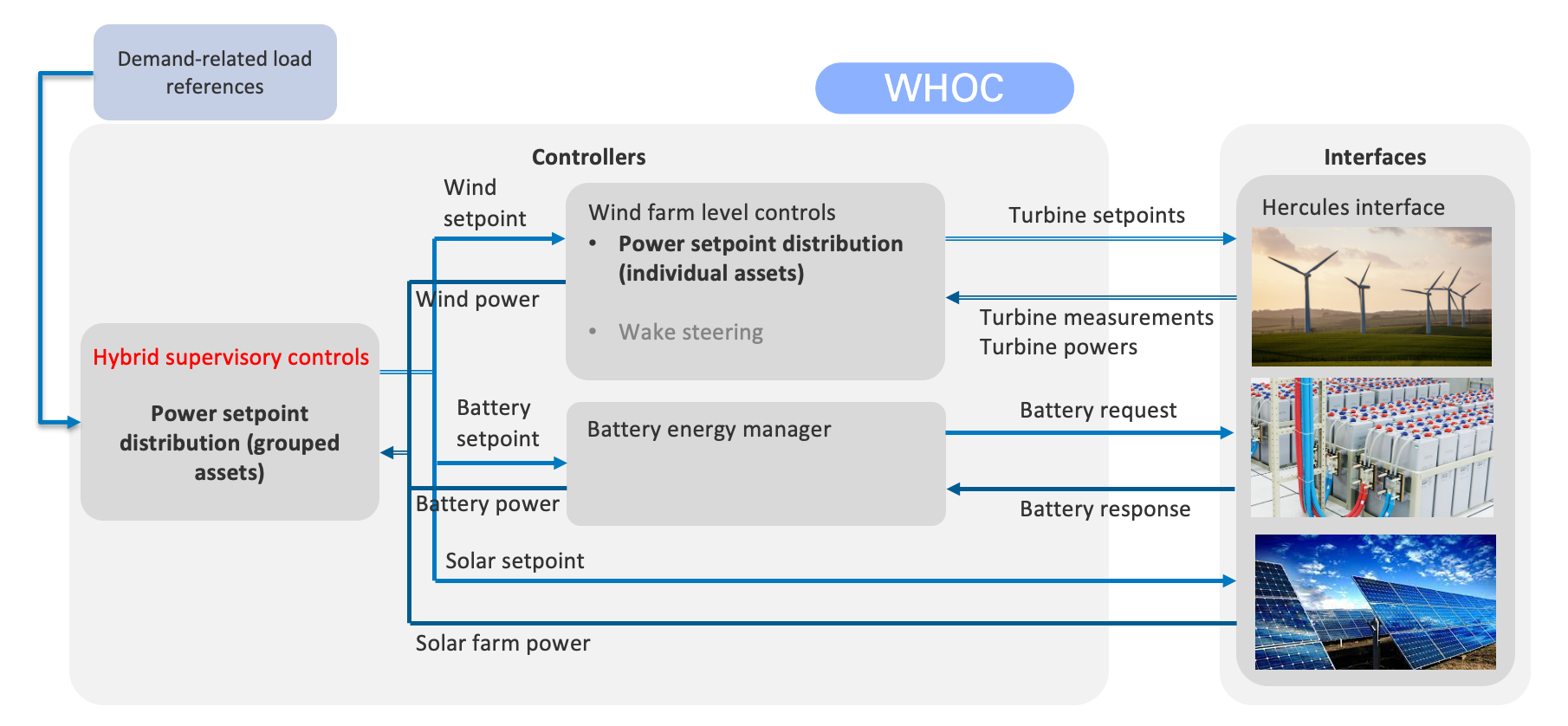}
    \caption{Overview of the Wind Hybrid Open Controller (WHOC) implementation with the proposed hybrid supervisory controller (at higher level), and low-level controllers for individual plants.}
    \label{fig:whoc_big_picture}
\end{figure}

The primary objective of the hybrid power plant supervisory controller is to use and coordinate its components (the wind farm, solar farm, and battery system) to meet power demands, subject to operational constraints, including respecting power limits of each component. A main difficulty in achieving this primary objective for a hybrid power plant is the lack of precise weather forecasting and modeling to anticipate the maximum available power that can be provided from the wind and sun. Here, we focus on achieving this primary objective despite uncertainties in the weather forecasting by leveraging feedback optimization (see Section \ref{ssec:fo} for background on feedback optimization). Figure \ref{fig:whoc_big_picture} shows the overall architecture of WHOC   where the proposed supervisory feedback optimization controller sends the optimized power set points to the low-level controllers for individual renewable and storage plants. 

To begin, consider the three components of the hybrid plant: wind, solar, and battery (storage). Each of these components has their own dedicated low-level controls to be able to track a given power set point, assuming the components have sufficient power to provide. To fit within the feedback optimization framework, we assume the supervisory control then operates at a slower timescale than the low-level controllers. This allows us to model the subsystems of wind, solar, and battery as input-output mappings, wherein the inputs are the power set points and the outputs are the true power. 
Let $P_w, P_s, P_b \in \mathbb{R}$ be the output power provided by the wind, solar, and battery components, respectively. Let $P_{w_d}, P_{s_d}, P_{b_d}\in \mathbb{R}$ be the input power set points provided to the wind, solar, and battery components, respectively. Then we can define the input-output mappings as:

\begin{subequations}\label{eq:subsytem input-output maps}
    \begin{equation}
    P_w = h_w(P_{w_d}, \myvar{d}_w)
\end{equation}
\begin{equation}
    P_s = h_s(P_{s_d}, \myvar{d}_s)
\end{equation}
\begin{equation}
    P_b = h_b(P_{b_d}, \myvar{d}_b)
\end{equation}
\end{subequations}
where $h_w: \mathbb{R}\times \mathbb{R}^{n_{d_w}} \to \mathbb{R}$, $h_s: \mathbb{R}\times \mathbb{R}^{n_{d_s}} \to \mathbb{R}$, $h_b: \mathbb{R}\times \mathbb{R}^{n_{d_b}} \to \mathbb{R}$ are the respective input-output mappings for the wind, solar, and battery components. The terms $\myvar{d}_{w} \in \mathbb{R}^{n_{d_w}}, \myvar{d}_s \in \mathbb{R}^{n_{d_s}}, \myvar{d}_b \in \mathbb{R}^{n_{d_b}} $ are the uncertain disturbances for the wind, solar, and battery components, respectively. For example, wind farms are affected by wake disturbances depending on the downstream position of the individual turbines with respect to one another. Another possible disturbance is uncertain wind gusts or wind behavior that may differ from predicted wind due to inherent difficulties in modeling wind. Such disturbances can be compensated for, assuming they are either constant or change slowly with respect to the timescale of the feedback optimization control. For solar plants, unexpected cloud cover or other impediments to the sun's rays would contribute to the unknown disturbance $\myvar{d}_s$. The battery component may have fewer disturbances associated with ambient effects yet may still be affected by modeling errors.

Let $P_{max_w}, P_{max_s}, P_{max_b} \in \mathbb{R}_{>0}$ be the maximum rated power of the wind farm, solar farm, and battery respectively. Furthermore, let $P_{min_b} \in \mathbb{R}$ be the minimum power limit of the battery. The power set points sent to each plant component must respect the component's maximum power capabilities. This is written as follows:
\begin{subequations}
    \begin{equation}
    0 \leq P_{w_d} \leq P_{max_w}
\end{equation}
\begin{equation}
    0 \leq P_{s_d} \leq P_{max_s}
\end{equation}
\begin{equation}
    P_{min_b} \leq P_{b_d} \leq P_{max_b}
\end{equation}
\end{subequations}

In addition to meeting power demands, the hybrid plant is able to generate excess power that is sent to the battery as storage. To accommodate this, we will consider $P_{w_l}, P_{w_b} \in \mathbb{R}$ as the component of the wind farm power that is used to meet the power demand and sent to the battery, respectively. Similarly, $P_{s_l}, P_{s_b} \in \mathbb{R}$ represent the component of the solar farm power that is used to meet the power demand and sent to the battery. It is clear that $P_w = P_{w_l} + P_{w_b}$ and $P_s = P_{s_l} + P_{s_b}$. In other words the sum of the power sent to the battery and sent to meet the power demand is the total power produced for the wind farm and the solar farm. This allows us to separate the battery power into charging and discharging components as follows:
\begin{subequations}\label{eq:battyer discharging/charging}
\begin{equation}
    P_b = P_{b_l} \  \text{(Discharging)}
\end{equation}
\begin{equation}
    P_b = P_{b_c} = P_{w_b} + P_{s_b} \ \text{(Charging)}
\end{equation}
\end{subequations}
where $P_{b_l} \in \mathbb{R}$ is the power discharged from the battery to meet the power demand and $P_{b_c} \in \mathbb{R}$ is the power charging the battery which comes from the excess power of the wind farm and solar farm. 

A similar split in power is defined in the control set points, i.e., $P_{w_d} = P_{w_{d_l}} + P_{w_{d_b}}$, $P_{s_d} = P_{s_{d_l}} + P_{s_{d_b}}$, where $P_{w_{d_l}}, P_{s_{d_l}} \in \mathbb{R}_{>0}$ are the power set points from the respective wind and solar farms to be sent to meet the power demands, and $P_{w_{d_b}}, P_{s_{d_b}}\in \mathbb{R}_{\geq 0}$ are the respective power set points sent to charge the battery. We will denote $P_{b_{d_l}} \in \mathbb{R}$ as the power set point sent to discharge the battery to meet the power demands.

Most energy storage systems do not allow for simultaneous charging/discharging. To account for this, we consider the following constraints to be used either in charging mode or in discharging mode:
\begin{equation}\label{eq:constraint discharging}
    P_{w_{d_b}} + P_{s_{d_b}} = 0 \ \text{(Discharging)}
\end{equation}
\begin{equation}\label{eq:constraint charging}
    P_{b_{d_l}} = 0 \ \text{(Charging)}
\end{equation}
In Eq. \eqref{eq:constraint discharging}, the power set points for excess wind and solar are restricted to zero for discharging mode because the battery cannot charge the excess power. In Eq. \eqref{eq:constraint charging}, the power set point to discharge the battery is set to zero for charging mode.

Finally, recall that the primary objective is to meet the power demand. Other secondary objectives can include charging the battery if there is excess power available from the wind farm and solar farm. Alternatively, a secondary objective could instead be to reduce the operational cost of the hybrid plant by curtailing the wind/solar farms to avoid producing excess power. All these objectives can be combined into the cost function $\ell$. An example of such a cost function is:
\begin{equation}\label{eq:example cost}
    \ell(P_{w_l}, P_{s_l}, P_{b_l}) = Q_r \frac{1}{2}(P_{w_l} + P_{s_l} + P_{b_l} - P_r)^2 + Q_b P_{b_l}
\end{equation}
where $P_r \in \mathbb{R}_{>0}$ is the reference power demand that the supervisory control should provide and $Q_r, Q_b \in \mathbb{R}_{\geq 0}$ are the respective gains for the power demand tracking and battery usage. The first term in Eq. \eqref{eq:example cost} is the error between demanded and provided power. The second term is the power discharged by the battery to adjusti how much of the power is supplied by the battery. A low ratio of $Q_b$ to $Q_r$ means to avoid discharging the battery as much as possible, and vice versa. We can define the optimization problem:
\begin{subequations}\label{eq:problem formulation}
\begin{align}
  &  \underset{P_{w_{d_l}}, P_{s_{d_l}}, P_{b_{d_l}}, P_{w_{d_b}}, P_{s_{d_b}}} {\text{min}} \hspace{.3cm} \ell(P_{w_l}, P_{s_l}, P_{b_l})   &&\\
 & \hspace{0.7cm} \text{s.t. }  \nonumber&&\\
&  \hspace{1cm}  P_{w_l} = h_w(P_{w_{d_l}}, \myvar{d}_w) &&\\
&  \hspace{1cm}  P_{w_b} = h_w(P_{w_{d_b}}, \myvar{d}_w) &&\\
&  \hspace{1cm}  P_{s_l} = h_s(P_{s_{d_l}}, \myvar{d}_s)  &&\\
&  \hspace{1cm}  P_{s_b} = h_s(P_{s_{d_b}}, \myvar{d}_s)  &&\\
&  \hspace{1cm} P_b = h_b(P_{b_{d_l}}, \myvar{d}_b) \ \text{(During discharging)} &&\\
&  \hspace{1cm} P_b = h_b(P_{w_b} + P_{s_b}, \myvar{d}_b) \ \text{(During charging)} &&\\
&\hspace{1cm} P_{w_d} = P_{w_{d_l}} + P_{w_{d_b}} \leq P_{max_w} \label{eq:max wind}&&\\
&\hspace{1cm} P_{s_d} = P_{s_{d_l}} + P_{s_{d_b}} \leq P_{max_s} \label{eq:max solar}&&\\
&\hspace{1cm} P_{min_b} \leq P_{b_{d_l}} \leq P_{max_b}   \ \text{(During discharging)} &&\\
&\hspace{1cm} P_{min_b} \leq P_{w_{d_b}} + P_{s_{d_b}} \leq P_{max_b}   \ \text{(During charging)}&&\\
&\hspace{1cm}  0 \leq P_{w_{d_l}},  0 \leq P_{w_{d_b}}, 0 \leq P_{s_{d_l}},  0 \leq P_{s_{d_b}}, 0 \leq P_{b_{d_l}}  &&\\
&\hspace{1cm} P_{w_{d_b}} + P_{s_{d_b}} = 0 \ \text{(During discharging)} &&\\
&\hspace{1cm} P_{b_{d_l}} = 0 \ \text{(During charging})
\end{align}
\end{subequations}
It is important to note that the above optimization problem takes into account both possible modes of the hybrid plant wherein the battery is either discharging or charging. When in discharging mode, the charging constraints are inactive, and vice versa for charging mode.

\section{Methodology}\label{sec:solution}

The problem defined in Eq. \eqref{eq:problem formulation} specifies what the supervisory controller must solve at all times. Each time the problem~\eqref{eq:problem formulation} is solved, each set point, $P_{w_d}, P_{s_d}, P_{b_d}$, is sent to the wind farm, solar farm, and battery, respectively, and the respective power measurements $P_{w_l}, P_{w_b}, P_{s_l}, P_{s_b}, P_b$ are collected and sent to the supervisory control to recompute the solution, and so on. This \textit{feedback loop} is represented by the online feedback optimization block from Fig. \ref{fig:schema}. We note that here we do not consider the switching mechanism between charging and discharging of the battery. In future work, this switching will be further investigated. 
The problem in Eq. \eqref{eq:problem formulation} can be written as the problem solved by the feedback optimization in Eq. \eqref{eq:fo main problem} as follows: 

Let $\myvar{y}  = (P_{w_l}, P_{w_b}, P_{s_l}, P_{s_b}, P_b)$, $\myvar{u}$ $=$ $(P_{w_{d_l}}$ $, $ $P_{s_{d_l}}$ $, $ $P_{b_{d_l}}$ $, $ $P_{w_{d_b}}$ $, $ $P_{s_{d_b}})$. The mapping $h$ is then the concatenation of all functions $h_w, h_s, h_b$ such that $\myvar{y} = h(\myvar{u}, \myvar{d})$, where $\myvar{d}$ is the concatenation of all disturbances $\myvar{d}_w, \myvar{d}_s, \myvar{d}_b$. Then $g$ is the concatenation of all the inequality constraints, and $c$ is the concatenation of all the equality constraints.
Note that in this case the problem \eqref{eq:fo main problem} is simplified since $g$ and $c$ are only a function of $\myvar{u}$, i.e., $g(\myvar{u}) \leq 0, c(\myvar{u}) = 0$, which implies that $\nabla_y g \equiv 0, \nabla_y c \equiv 0$. Similarly, $\ell$ is only a function of the output $\myvar{y}$, i.e., $\ell(\myvar{y})$. Under this formulation, problem \eqref{eq:problem formulation} can be solved either in discharging or charging mode using the controller \eqref{eq:FO control}. For clarity we rewrite the simplified form of problem \eqref{eq:fo main problem} as follows:
\begin{subequations}\label{eq:hybrids main problem}
\begin{align}
 \myset{P}_h: \hspace{0.4cm} &  \underset{\myvar{u}} {\text{min}} \hspace{.3cm} \ell \left(h(\myvar{u}, \myvar{d}) \right)   &&\\
 & \hspace{0.1cm} \text{s.t. }  \ g( \myvar{u}) \leq 0 &&\\
  & \hspace{0.85cm}  c( \myvar{u}) = 0
\end{align}
\end{subequations}

The feedback optimization controller applied to $\myset{P}_h$ can be simplified to:
\begin{equation}\label{eq:hybrid FO control}
    \dot{\myvar{u}}(t) = \eta F(\myvar{y}(t), \myvar{u}(t)), \ \myvar{u}(0) = \myvar{u}_0
\end{equation}
\begin{subequations}\label{eq:hybrid fo F optimization}
\begin{align}
 F(\myvar{y}, \myvar{u}) = \hspace{0.4cm} &  \underset{\myvar{\theta}} {\text{argmin}} \hspace{.3cm} \|\myvar{\theta} + \nabla_u h(\myvar{u})^T \nabla_y \ell(\myvar{y})^T  \|_2^2   &&\\
 & \hspace{0.1cm} \text{s.t. }  \nonumber&&\\
&  \hspace{1.4cm}  \nabla_u g( \myvar{u})^T \myvar{\theta} \leq -\beta g( \myvar{u}) &&\\
&  \hspace{1.4cm}  \nabla_u c( \myvar{u})^T \myvar{\theta} = -\beta c( \myvar{u}) &&\\
&  \hspace{1.4cm} w(\myvar{\theta}) \leq 0 \label{eq:rate limit}
\end{align}
\end{subequations}
The rate limit constraint $w$ can be used to limit power rate terms in the wind farm, solar farm, and/or battery storage. In this problem formulation $\myvar{\theta}$ represents the derivative of the control set points, i.e., $\myvar{\theta} $$= $$(\dot{P}_{w_{d_l}}$ $, $ $\dot{P}_{s_{d_l}}$ $, $ $\dot{P}_{b_{d_l}}$ $, $ $\dot{P}_{w_{d_b}}$ $, $ $\dot{P}_{s_{d_b}})$.

In order to implement Eq. \eqref{eq:hybrid FO control}, we require the \textit{gradient} terms of all the input-output mappings and the constraint functions. 
We also require more explicit power limits to account for the fact that the wind and solar farms can only provide power that is less than or equal to the maximum available power depending on the current weather. By using real-time measurements of wind speed and irradiance to estimate the maximum available power for the wind and solar farms, the supervisory control can more effectively balance the use of wind/solar power with battery storage. In the following section, we present the gradient terms and the respective modeling of the maximum available power to implement the feedback optimization controller.

\subsection{Resource-Specific Considerations}\label{ssec:control modeling}

Here, we consider how the hybrid power plant supervisory control can be formulated as a feedback optimization control from Section \ref{ssec:fo}. The feedback optimization control has several advantages. It does not require prior knowledge of the  $h$ mapping or an estimate of $\myvar{d}$. Instead, the `model' in this case refers to the terms $\nabla_u h$, which effectively define the sensitivity of the output function with respect to the input. Furthermore, we show how real-time measurements can be used to determine available wind and solar power to aid the feedback optimization controller. In the following section, we show how to define the elements of the feedback optimization problem so that it can be applied as a hybrid power plant supervisory control law.

We begin with the input-output sensitivities defined by $\nabla_u h$. We define this sensitivity as simply the identity matrix:
\begin{equation}\label{eq:input-output hybrid sensitivity}
    \nabla_u h = I
\end{equation}

This implies that when a reference power set point is sent to each hybrid plant component, we expect the plant to output the required power. Of course this would only occur when there is enough energy from the wind/sun or in the battery to provide the necessary power. 

In order to address the case when there is insufficient power either from the wind/sun or stored energy, we restrict the allowable reference set points so that the reference power set points always remain feasible. In the next sections we define the appropriate restrictions of the power set points per individual hybrid plant component in order to ensure the sensitivity condition in Eq.\eqref{eq:input-output hybrid sensitivity} holds. We note that the modeling approach considered here is simple and relies on model approximations because the feedback optimization controller provides robustness to model errors by means of feedback measurements. This distinguishes the use of approximate models in our feedback optimization approach compared to existing open-loop dispatch schedulers. 

\subsubsection{Constraints on the Wind Subsystem}

To implement the feedback optimization control law for the hybrid plant, our next task is to define the restrictions on the wind power set points to ensure Eq. \eqref{eq:input-output hybrid sensitivity} holds. For the wind power plant, the power restriction is dependent on the wind speed.

If the wind speed is too low, there would be insufficient power to match a given power set point. Herein, we exploit existing models of the available wind power. The maximum available power of a wind turbine can be determined by the local wind speed $\omega_i \in \mathbb{R}_{\geq 0}$, air density $\rho \in \mathbb{R}_{>0}$, area swept by the rotor $A\in \mathbb{R}_{\geq  0}$, and power coefficient $C_p(\omega_i) \in \mathbb{R}_{\geq 0}$ \citep{Burton2002Wind}:
\begin{equation}\label{eq:available wind power}
    \bar{P}_{w_i}(\omega_i) = \max(0,  \frac{1}{2} \rho A C_p(\omega_i) \omega_i^3 )
\end{equation}
where $\bar{P}_{w_i}$ denotes the maximum available power of wind turbine $i$. Thus, for the entire wind farm, we can compute the total maximum available wind power as:
\begin{equation}\label{eq:total available wind power}
    \bar{P}_w(\myvar{\omega}) = \sum_{i=1}^{N_t} \bar{P}_{w_i}(\omega_i)
\end{equation}
where $\bar{P}_w$ is the total available wind power over all $N_t$ turbines and $\myvar{\omega} = [\omega_1, \omega_2, ..., \omega_{N_t}]$ is the collection of all local wind speeds. Note, the local wind speeds can be difficult to measure, so this equation can be approximated by assuming all local wind speeds are equal to the ambient wind speed. 

Now, we can define our constraint on the wind subsystem to ensure the control, i.e., the wind power set point, never exceeds the available wind power:
\begin{equation}\label{eq:max wind constraint}
 P_{w_d} = P_{w_{d_l}} + P_{w_{d_b}} \leq \min \{ P_{max_w}, \bar{P}_w(\myvar{\omega}) \}
\end{equation}
where $P_{max_w}$ is the maximum rated power of the wind farm. This max rated power ensures the wind farm remains within its operating regions for safety and maintenance purposes. Note that when the wind power curve reaches zero, the power set point is fixed to $P_{w_d} = 0$ so that the feedback optimization will never ask for power when there is insufficient power in the wind. In the problem formulation, Eq. \eqref{eq:max wind constraint} is substituted for Eq. \eqref{eq:max wind}.

\begin{figure}[t!]
    \centering
    \includegraphics[width=0.75\linewidth]{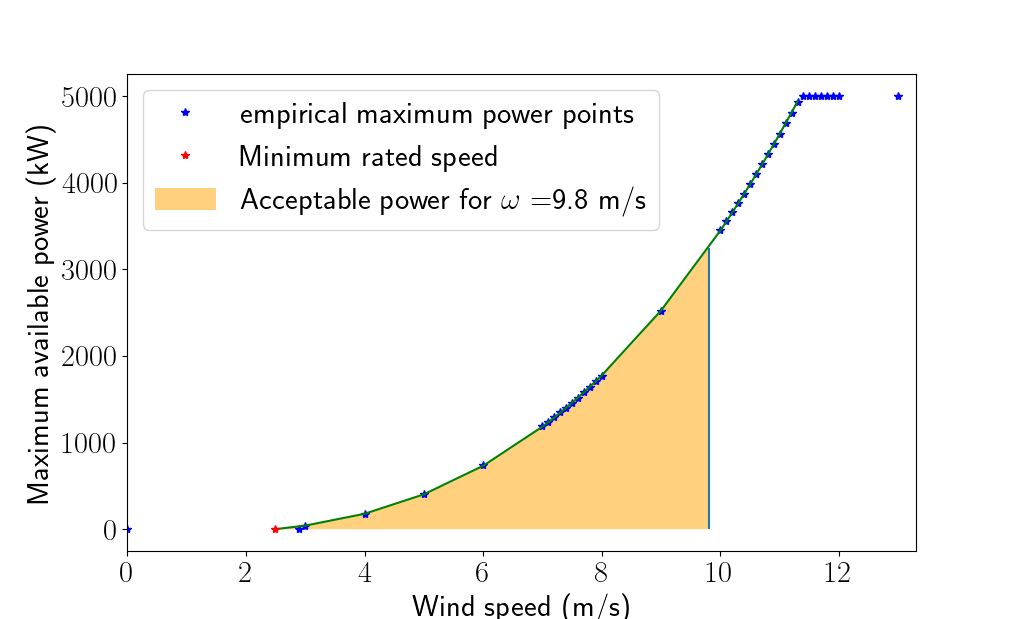}
    \caption{Example: Maximum available power for 5 MW turbine.}
    \label{fig:max power}
\end{figure}

For example, consider a 5 MW turbine. The empirical maximum power set points can be provided by wind control engineers, as shown in Fig. \ref{fig:max power}. A cubic polynomial curve can be fit to this to define $\bar{P}_w$, represented by the green curve in the figure. To determine the set of power set points that are allowed for a given speed, consider the case where the current wind speed is 9.8 m/s. In Fig. \ref{fig:max power}, the area in yellow represents the constraint region that the power set points must remain inside. In other words, the yellow region represents values of the power set point that satisfy \eqref{eq:max wind constraint}. Note that for a different wind speed, the right border of this region will shift appropriately and be upper-bounded by the curve defined by $\bar{P}_w$ and the maximum rated power of $P_{max_w} = 5000$ kW.

\subsubsection{Constraints on the Solar Subsystem}

Here, we define the restrictions of the reference power set points for the solar plant. We use a model to approximate the maximum available power that a solar plant can provide. This model depends on the total irradiance, denoted as $I_T\in \mathbb{R}$, which is determined according to \citep{deshmukh2008modeling}:
\begin{equation}
    I_{T} = I_b R_b + I_d R_d + (I_b + I_d) R_r
\end{equation}
where $I_b \in \mathbb{R}$ represents direct normal radiation, $I_d\in \mathbb{R}$ is for diffuse solar radiation, and corresponding tilt factors are denoted by $R_d$ (diffused) and $R_r$(reflected). Thereafter, the maximum possible power generated by the solar array can be calculated as follows:
\begin{equation}
    \bar{P}_{s}(I_t, T_{air}) = I_T \eta(T_{air}) A_{pv}
\end{equation}
where $\eta \in \mathbb{R}$ denotes the efficiency of the solar panel, which is a function of the ambient air temperature, $T_{air}$, and $A_{pv} \in \mathbb{R}_{>0}$ is the total area covered by the solar panels. 

\begin{figure}[H]
    \centering
    \includegraphics[width=0.75\linewidth]{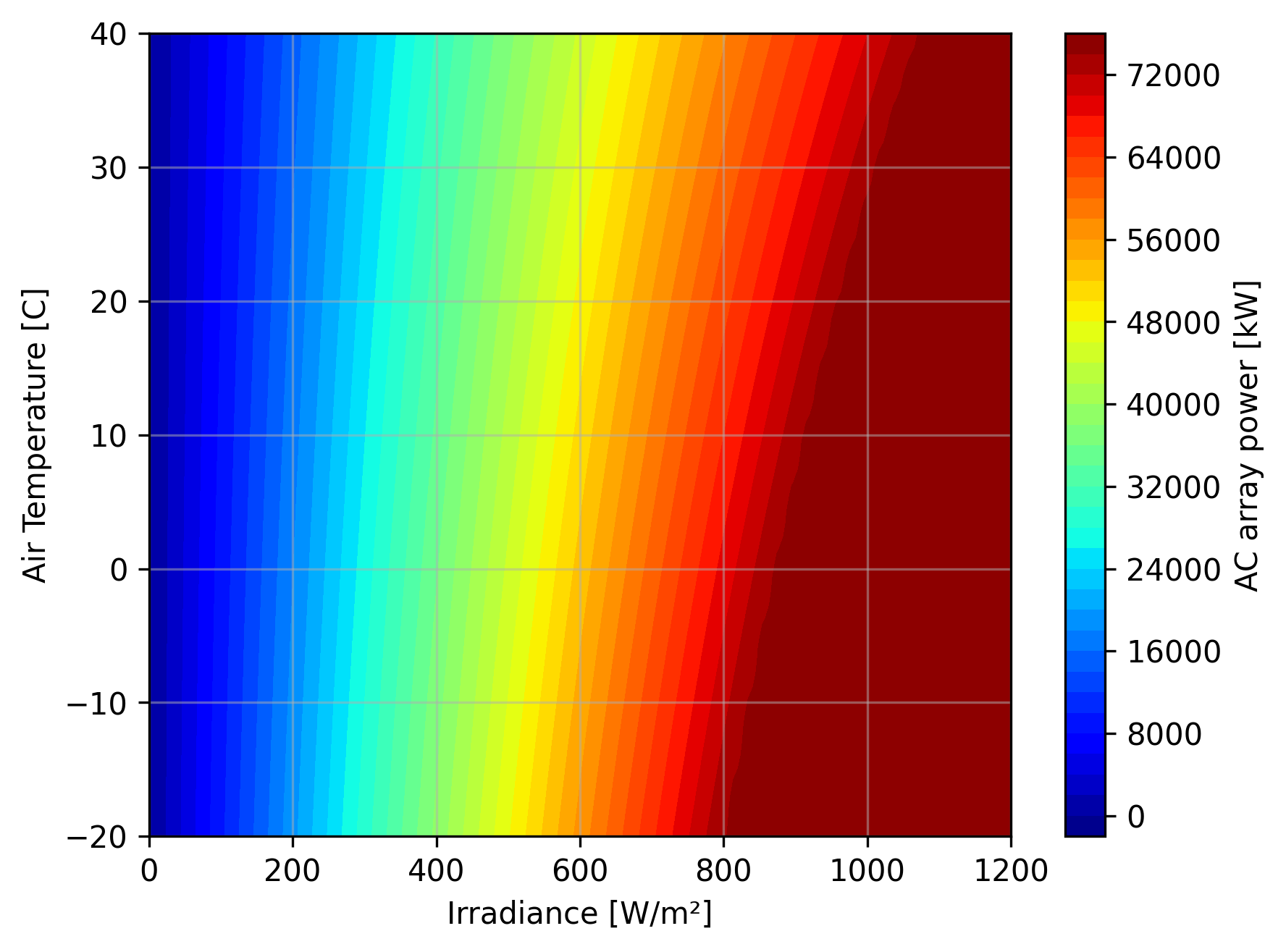}
    \caption{An example showing the maximum available power profile based on irradiance and air temperature generated using the National Renewable Energy Laboratory's (NREL's) System Advisor Model\texttrademark ~\citep{pysam_24}.}
    \label{fig:solar_fig}
\end{figure}

Figure \ref{fig:solar_fig} depicts the maximum available solar power based on the  irradiance and air temperature for a $100$ MW rated plant with a single-axis tracker.  

Thus, the constraint of the reference power set point for the solar farm is given as:
\begin{equation}\label{eq:max solar constraint}
P_{s_d} = P_{s_{d_l}} + P_{s_{d_b}} \leq \min \{ P_{max_s} \bar{P}_{s}(I_T,T_{air}) \},
\end{equation}
where $P_{max_s} \in \mathbb{R}_{>0}$ is the maximum rated power of the solar farm. This constraint, Eq. \eqref{eq:max solar constraint}, will replace Eq.  \eqref{eq:max solar}.

\subsubsection{Constraints on the Battery Energy Storage Subsystem}

In our approach, the battery undergoes a charging and discharging cycle in a coordinated manner with wind and solar plants. Following the Hercules implementation, we have categorized the following initial settings for the battery.
Let us denote the energy capacity  by $C$ in kWh, initial state of charge (SOC) by $S$, and maximum and minimum SOCs by $S_{max}$ and $S_{min}$. The charge (energy) limits are bounded by  $E_{min}$ =  $S_{min}C$, and $E_{max}$ = $S_{max}C$. 
The ramp up/down limits in kW/s are bounded by $R_{min}, R_{max}$. 
The current battery charge state is given by, $E = S C $.
The battery will feature a low-level controller responsible for adhering to charging and energy limitations. Let $P_{avail}$ represent the available charging power in kW, and let $P_{b_d}$ be the storage power reference. The present charging/discharging power is indicated by $P_{b}$ in kW. 

For the time step $\Delta T$, upper constraints are given by:
    \begin{align}
    \label{upper_cons}
        & c_{h1} = (E_{max} - E) / \Delta T, 
        c_{h2} = P_{max_b}, 
        c_{h3} = P_{avail}
    \end{align}
to ensure the battery does not exceed the maximum energy bound, respects the maximum rated power, and does not discharge more power than is available.
The lower constraints are given by: 
    \begin{align}
    \label{lower_cons}
        &c_{l1} = (E_{min} - E) / \Delta T, 
        c_{l2} = P_{min_b}  
    \end{align}
to ensure the battery does not exceed the minimum energy bound and respects the lower power limit.

If the reference commands are within the lower and upper constraints, then the battery is able to follow the reference and support the hybrid plant. 

Therefore, the constraint to the feedback optimization algorithm would restrict the reference power set point for the energy storage system as follows for both discharging and charging modes:
\begin{subequations}\label{eq:battery constraints}
    \begin{align}
 max([c_{l1}, c_{l2}]) \leq P_{b_{d_l}} \leq  min([c_{h1}, c_{h2}, c_{h3}]) \ \text{(Discharging)},
\end{align}
\begin{align}
 max([c_{l1}, c_{l2}]) \leq P_{w_{d_b}} + P_{s_{d_b}} \leq  min([c_{h1}, c_{h2}])  \ \text{(Charging)}
\end{align}
\end{subequations}
The updated energy state is given by
\begin{align}
    E(t+\Delta T) = E(t ) + P_{b} \Delta T
\end{align}

\subsection{Simulation Tests with Realistic Resource Inputs and Simplified Models}

\begin{figure}[t!]
    \centering
    \begin{subfigure}[t]{0.45\linewidth}
        \centering
        \includegraphics[width=\linewidth]{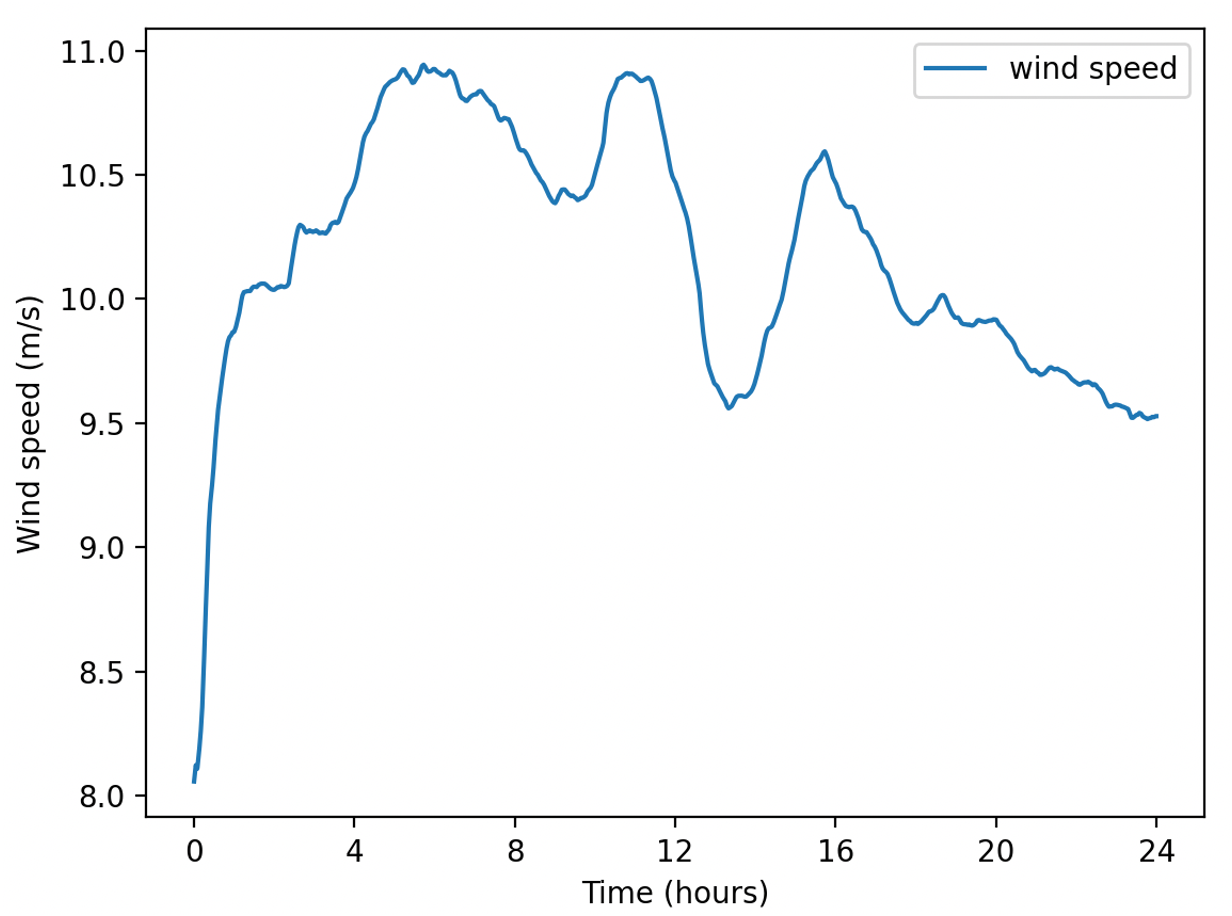}
        \caption{Realistic wind profile.}
        \label{fig:realistic_wind}
    \end{subfigure}
    \hfill
    \begin{subfigure}[t]{0.45\linewidth}
        \centering
        \includegraphics[width=\linewidth]{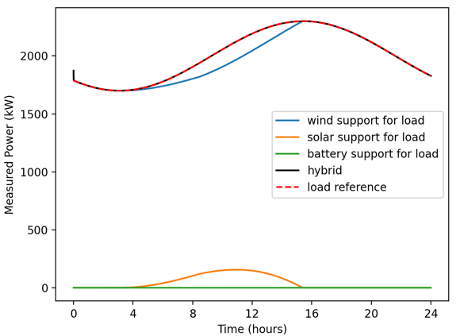}
        \caption{Supervisory control under moderate load (battery charging).}
        \label{fig:realistic_charging1}
    \end{subfigure}

    \vspace{1em} 

    \begin{subfigure}[t]{0.45\linewidth}
        \centering
        \includegraphics[width=\linewidth]{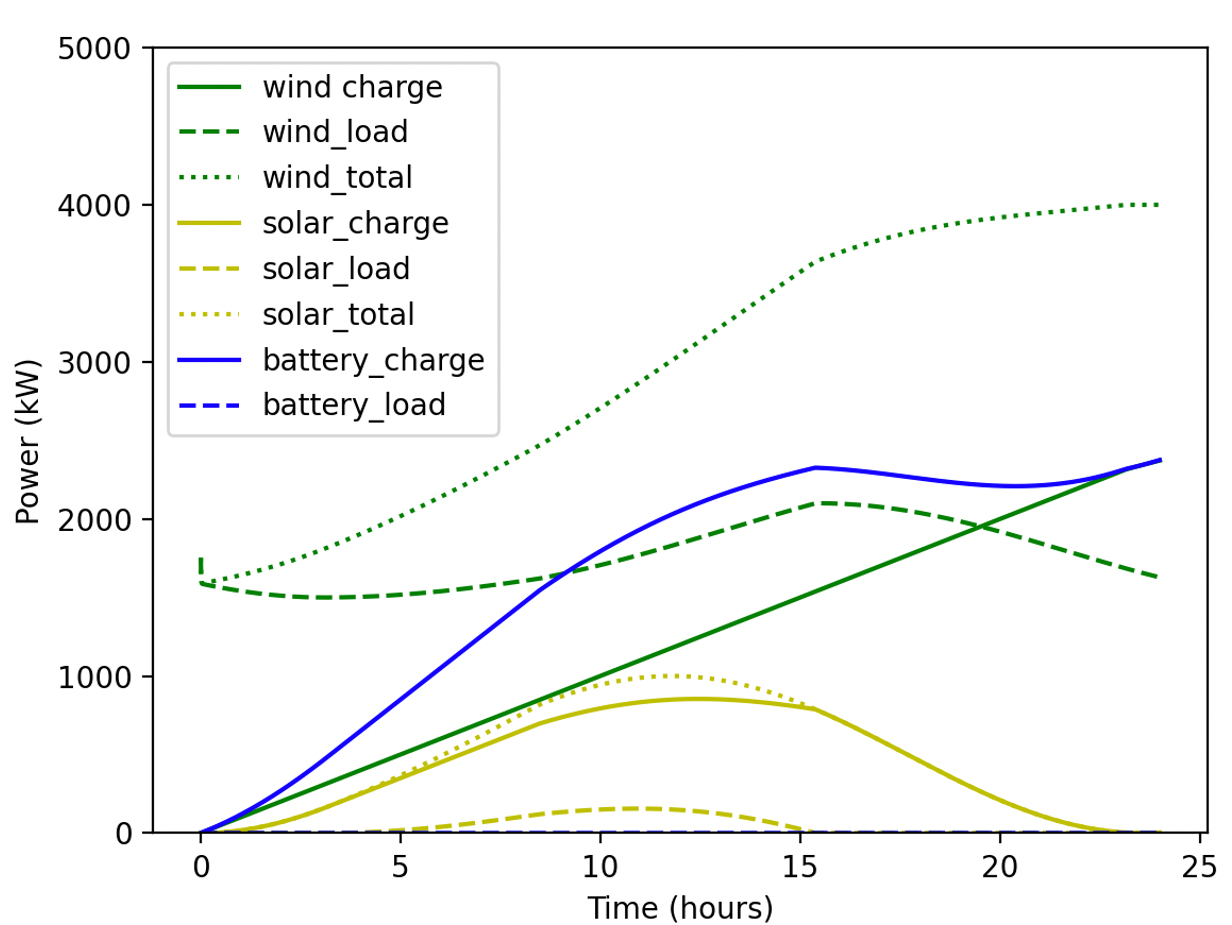}
        \caption{Component-wise load support showing battery charging from excess generation.}
        \label{fig:realistic_charging2}
    \end{subfigure}
    \hfill
    \begin{subfigure}[t]{0.45\linewidth}
        \centering
        \includegraphics[width=\linewidth]{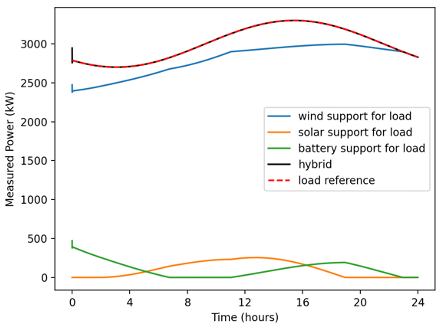}
        \caption{Battery discharging mode during high demand period.}
        \label{fig:realistic_discharging2}
    \end{subfigure}

    \caption{Hybrid plant operation scenarios under realistic conditions showing wind variability, supervisory control response, battery charging, and discharging behaviors.}
    \label{fig:realistic_2x2}
\end{figure}

This subsection outlines the controller performance with realistic scenarios for validating the supervisory control of hybrid power plants with simplified models. By integrating wind, solar, and power demand profiles into the feedback optimization computations, the aim is to enhance the accuracy and reliability of the control systems. Each hybrid plant component is modeled as a simple input-output model where the output power is equal to the reference set point power subject to the restrictions defined by Eqs. \eqref{eq:max wind constraint} and \eqref{eq:max solar constraint}.

Figure \ref{fig:realistic_wind} shows a realistic wind speed profile over a day, where wind speed fluctuates between 8.0 m/s and 11.0 m/s to simulate how the wind power contributes to the overall energy mix. During periods of moderate power demand, excess wind and solar power can be utilized to charge the battery, ensuring efficient energy storage and utilization. Figure \ref{fig:realistic_charging1} shows that during moderate demand periods, wind and solar energy are sufficient, and the battery discharges only when necessary. The hybrid system's output closely aligns with the load reference, with any excess power directed towards charging the battery. Figure \ref{fig:realistic_charging2} provides details on the component power sharing, highlighting the roles of wind and solar power in battery charging.

Figure \ref{fig:realistic_discharging2} represents a scenario with increased power demand, showing how the hybrid system reacts. The plot shows that wind power alone is insufficient during high demand periods. Consequently, battery discharging becomes necessary to meet the load. During these periods, both wind and solar power contribute, but the battery also discharges to supply the needed power. The approach ensures that hybrid systems can manage power generation and storage dynamically, meeting demand reliably and efficiently.

\section{Supervisory Controller Integration with a Co-simulation Hybrid Plant Engine and Validations}
\label{sec:high_fidelity}

\begin{figure}[h]
    \centering
    \includegraphics[width=0.9\linewidth]{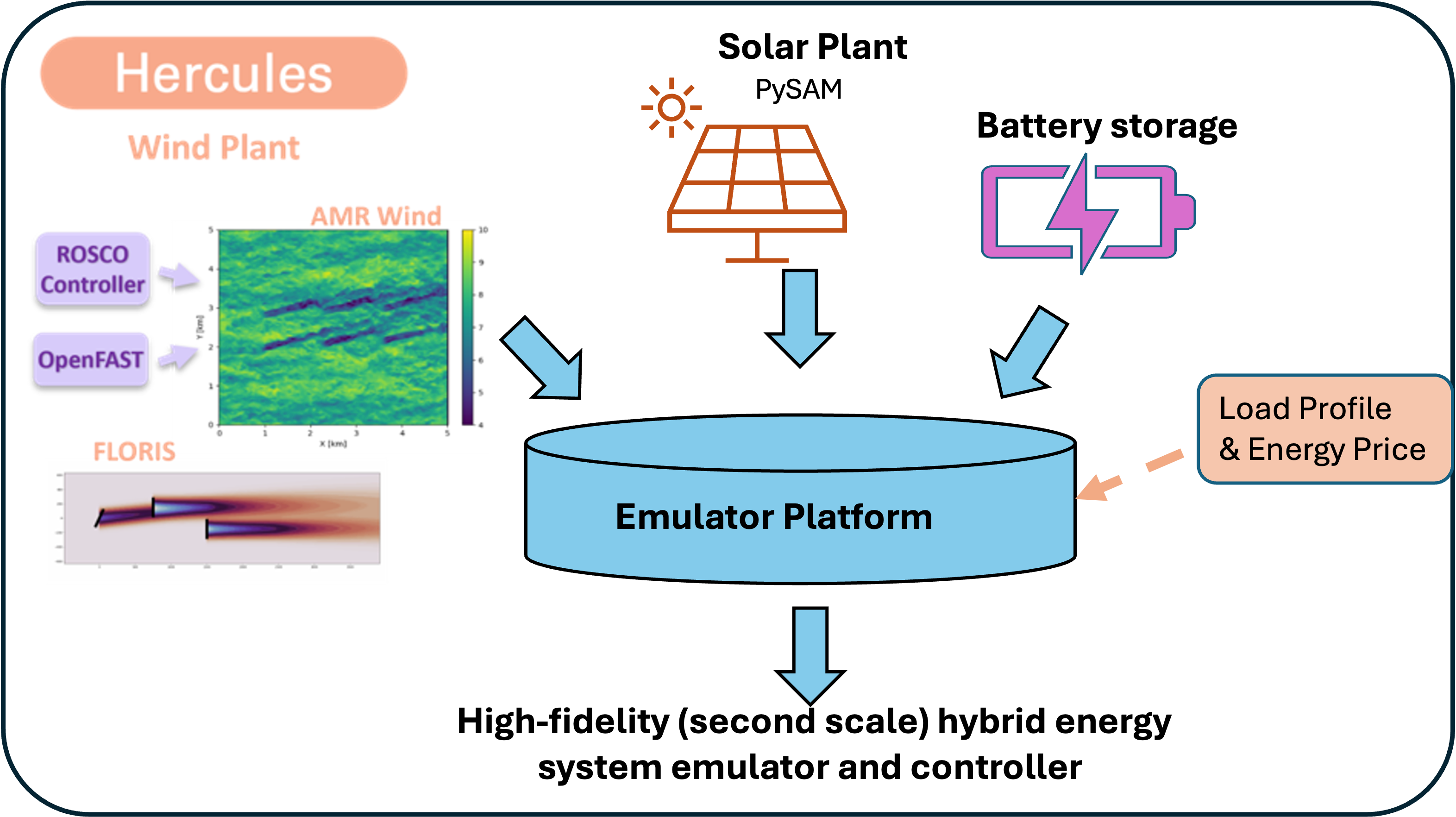}
    \caption{Schematic of the Hercules co-simulation platform.}
    \label{fig: hercules}
\end{figure}

\begin{figure}[]
    \centering
    \includegraphics[width = \linewidth]{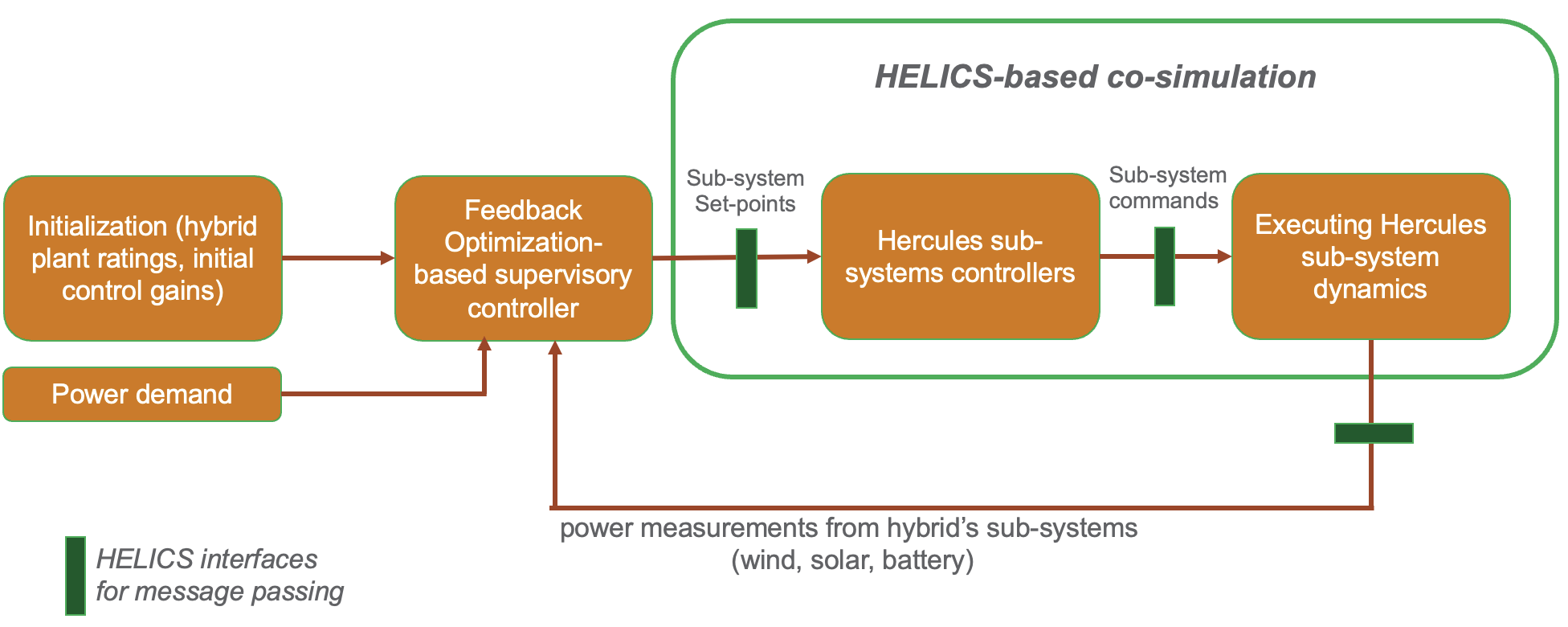}
    \caption{Flowchart for Hercules integration.
}
    \label{fig:hercules_flow}
\end{figure}

The Hercules simulation platform \citep{Hercules_2024} was utilized to co-simulate these hybrid plant models. Designed specifically to support the concurrent simulation of diverse technologies, this platform facilitates both the development and validation of control algorithms, as illustrated in Fig. \ref{fig: hercules}. To synchronize the timing of these technologies, Hercules integrates the HELICS software module \cite{hardy2024helics}.
Figure \ref{fig:hercules_flow} illustrates how the supervisory control is integrated within the Hercules framework, utilizing HELICS Co-Simulation for communication in hybrid power plants. In this flexible framework, the blocks in Fig. \ref{fig:hercules_flow} can easily be exchanged with different models, enabling the advanced feedback optimization hybrid controller to replace an existing rule-based controller block. Within the Hercules framework, multiple component-level controllers work together to manage the entire hybrid power plant. The internal subsystem control handles low-level controllers for each subsystem, while the Hybrid Supervisory Controller is the module where feedback optimization-based supervisory control is implemented. These routines are part of the WHOC module \citep{whoc-docs}.

The feedback optimization hybrid controller begins by loading an input dictionary file, which contains key parameters and configurations for the controller. This dictionary is then used to configure the hybrid controller through the Hybrid Controller function definitions. Figure \ref{fig:hercules_flow} shows the full integration workflow and where  supervisory controller is integrated.

\subsection{Numerical Experiments}
\label{subsec:numericals}

\begin{table}[t!]
\centering
\caption{Hybrid Power Plant Controller Configuration}
\label{tab:controller_config}
\renewcommand{\arraystretch}{1.2}
\begin{tabular}{llr}
\hline
\textbf{Parameter} & \textbf{Description} & \textbf{Value} \\
\hline
\multicolumn{3}{l}{\textbf{Controller Settings}} \\
\hline
num\_turbines & Number of wind turbines & 10 \\
wind\_capacity\_MW & Rated wind capacity & 50 MW \\
solar\_capacity\_MW & Rated solar capacity & 100 MW \\
battery\_capacity\_MW & Rated battery capacity & 20 MW \\
\hline
\multicolumn{3}{l}{\textbf{Hybrid Controller Settings}} \\
\hline
dt & Controller time step & 0.03 s \\
components & Active components & wind, solar, battery \\
battery\_mode & Battery operational mode & discharge \\
\hline
\multicolumn{3}{l}{\textbf{Controller Gains}} \\
\hline
$\eta$ & Control gain ($\dot{u} = \eta F$) & 0.95 \\
$\beta$ & Constraint gain ($\dot{h} \leq \beta h$) & 1.0 \\
\hline
\multicolumn{3}{l}{\textbf{Reference Limits (kW)}} \\
\hline
$P_{max_w}$ & Max wind setpoint & 50000 \\
$P_{max_s}$ & Max solar setpoint & 100000 \\
$P_{max_b}$ & Max battery setpoint & 20000 \\
$P_{min_w}$ & Min wind setpoint & 0 \\
$P_{min_s}$ & Min solar setpoint & 0 \\
$P_{min_b}$ & Min battery setpoint & -20000 \\
\hline
\end{tabular}
\end{table}

We have subsequently integrated the feedback-optimization-based supervisory controller with the Hercules simulation platform for a representative hybrid power plant. Due to computational complexity with simulations, we showcase the controller performance for a few minutes of simulations. The plant's wind and solar generation resources are rated at $50$ MW and $100$ MW, respectively. The plant battery is a 4-hour, $20$ MW battery. The solar farm is composed of single-axis solar panels.
We considered $10$ turbines for the wind farm, each rated at $5$ MW. The turbines are laid out in a grid fashion in three rows: the first row has four turbines, and the second and third rows each have three turbines. The NREL 5 MW reference design is used for the turbines, with a diameter of $126$ m \citep{Jonkman09-5mw}. The wakes are simulated with a steady-state wake model in the FLORIS simulator \citep{floris-docs}. Simulated wind resource data are used in the study by perturbing the mean wind speed at every 10 minutes by uniform random samples in the range $\pm 2$ m/s. We considered mean  wind speeds of $8$ m/s and $9.5$ m/s for the charging and discharging conditions, respectively, and uniformly generated samples within the $\pm 5\%$ range, and then performing spline based fitting.  

In this simulation, we use a variable wind speed, but we keep the wind direction constant. The solar farm is simulated in the PySAM module, with the physical location at latitude: 39.7442$^{\circ}$, longitude: -105.1778$^{\circ}$, elevation: 1829 m \citep{pysam_24}.   The solar input was drawn from the National Solar Radiation Database \citep{sengupta_national_2018} at the minute scale, which is the smallest time interval option. The solar data are then linearly interpolated between each data point as the input into the Hercules solar simulation model. The load tracking references are taken from PJM Interconnection, a regional transmission organization in the eastern United States. We used their \textit{REG-A} signal that provides regulation information over a period of time throughout the day. Table \ref{tab:controller_config} shows more details on the hybrid supervisory controller configuration.

\begin{figure}[t!]
    \centering
    \begin{subfigure}[t]{0.65\linewidth}
        \centering
        \includegraphics[width=\linewidth]{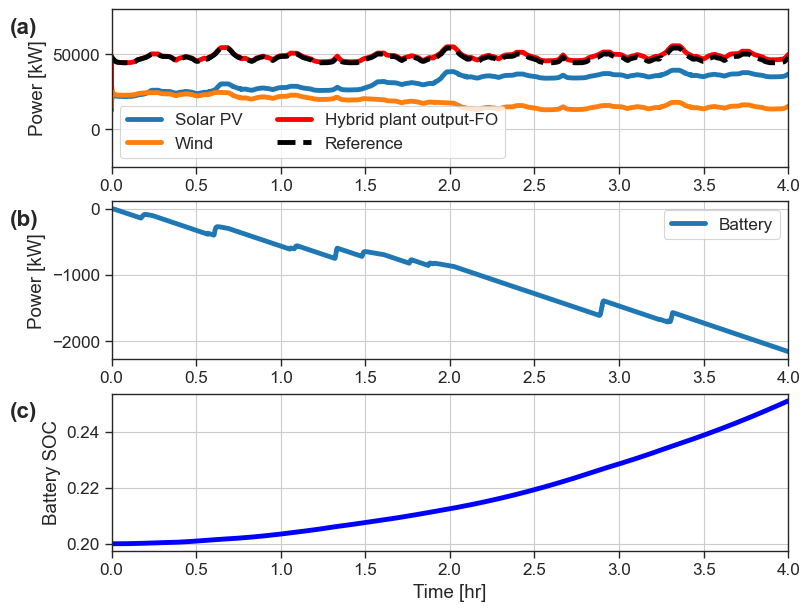}
        \caption{Hybrid power plant's subsystem outputs during charging mode.}
        \label{fig:output_Charging}
    \end{subfigure}
    \hfill
    \begin{subfigure}[t]{0.45\linewidth}
        \centering
        \includegraphics[width=\linewidth]{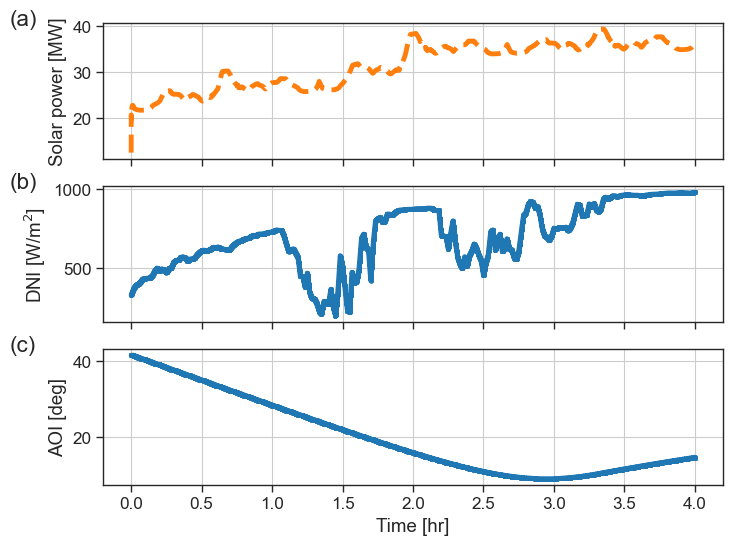}
        \caption{Solar plant outputs and resource availability during charging.}
        \label{fig:output_SolarProfile_Charging}
    \end{subfigure}
    \hfill
    \begin{subfigure}[t]{0.45\linewidth}
        \centering
        \includegraphics[width=\linewidth]{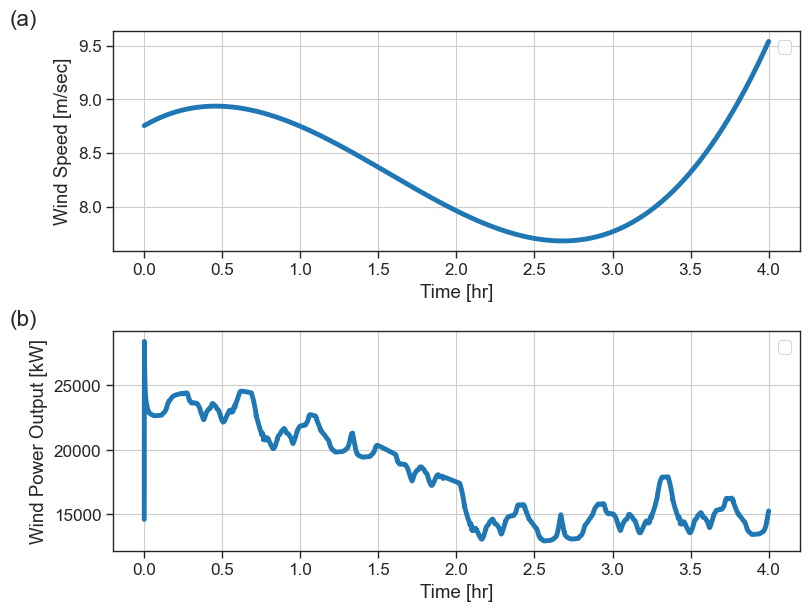}
        \caption{Wind plant outputs and resource availability during charging.}
        \label{fig:output_WindProfile_Charging}
    \end{subfigure}

    \caption{Hybrid power plant operation under battery charging conditions showing contributions from (a) overall plant, (b) solar subsystem, and (c) wind subsystem.}
    \label{fig:charging_3panel}
\end{figure}

\noindent \textit{Charging scenario with moderate demand:} In the charging mode, the plant is tasked with delivering close to $50$ MW of demand. Figure \ref{fig:output_Charging} shows the tracking capability of the plant in delivering the required power to meet the demand requirement. The battery charging profile shows steady increase during this process. The solar plant supports $30$ to $40$ MW based on the availability of resources, as shown in Fig. \ref{fig:output_SolarProfile_Charging} with the direct normal irradiance (DNI) reaches close to $1000$ W/m\textsuperscript{2}, and the angle of incidence (AOI) decreases over time. Figure \ref{fig:output_WindProfile_Charging} shows that the wind plant support decreases over time based on the wind speed, as shown in the figure and wind direction set at 164.88$^{\circ}$. For the charging mode, we considered specific objectives such as:
\begin{equation}\label{eq: cost charging}
    \ell(P_{w_l}, P_{s_l}, P_{b_l}) = Q_r \frac{1}{2}(P_{w_l} + P_{s_l} + P_{b_l} - P_r)^2 - Q_b (P_{w_b} + P_{s_b})
\end{equation}
where the first term makes sure that the total resource generation matches with the requested demand $P_r$, and the second term maximizes the charging power to the battery from wind ($P_{w_b}$) and solar ($P_{s_b}$). Gains are set at $Q_r = 45$, and $Q_b = 2$, after tuning for optimized trade-off performance. Additional constraints on the control actions are considered as:
\begin{align}
    P_{b_{d_l}} = 0,
\end{align}
that is, the control commands to send battery power to load is set to zero. In this scenario, the system has encountered sudden falls in solar DNI, and also slowly varying decrement of the wind speeds to test the robust behavior of the online control algorithm. This study shows that in presence of variable solar and wind conditions, the feedback optimization based real-time closed-loop controller can optimize the supervisory control set-points, and support the load, and concurrently charge the battery when resources are sufficient.

\begin{figure}[t!]
    \centering
    \begin{subfigure}[t]{0.65\linewidth}
        \centering
        \includegraphics[width=\linewidth]{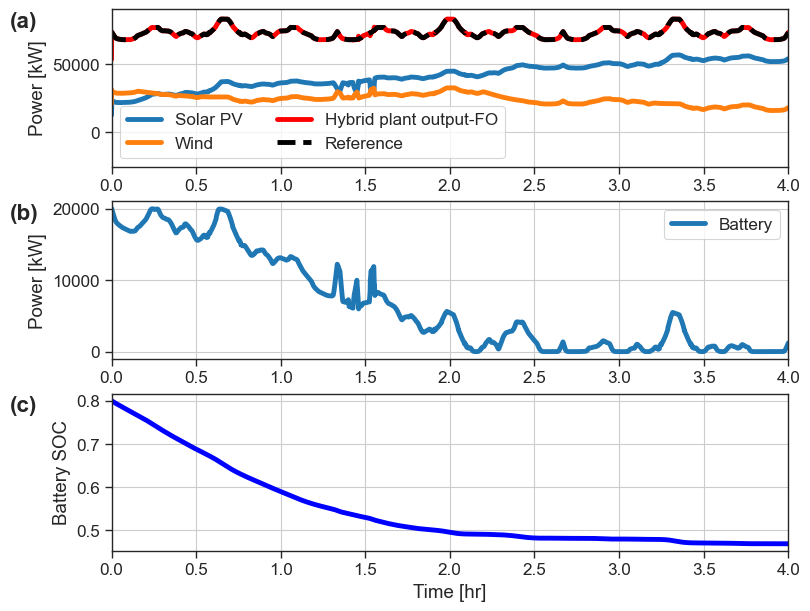}
        \caption{Hybrid power plant subsystem outputs during discharging mode.}
        \label{fig:output_discharging}
    \end{subfigure}

    \vspace{1em} 

    \begin{subfigure}[t]{0.45\linewidth}
        \centering
        \includegraphics[width=\linewidth]{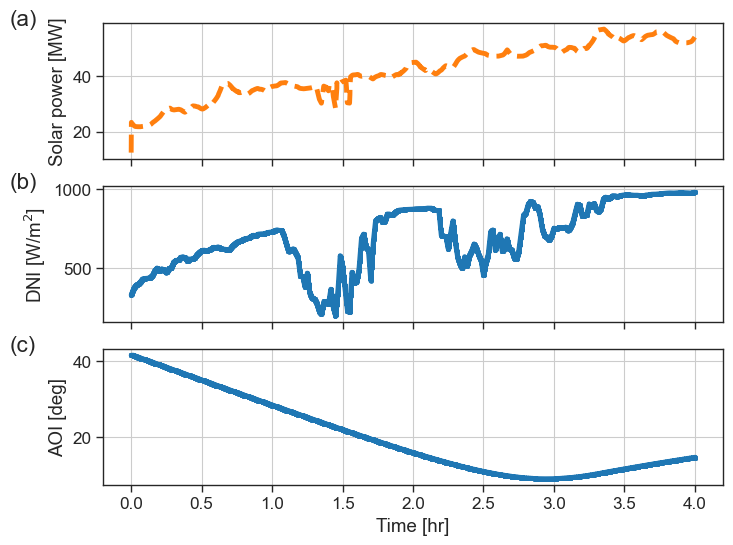}
        \caption{Solar plant outputs and resource availability during discharging.}
        \label{fig:output_SolarProfile_discharging}
    \end{subfigure}
    \hfill
    \begin{subfigure}[t]{0.45\linewidth}
        \centering
        \includegraphics[width=\linewidth]{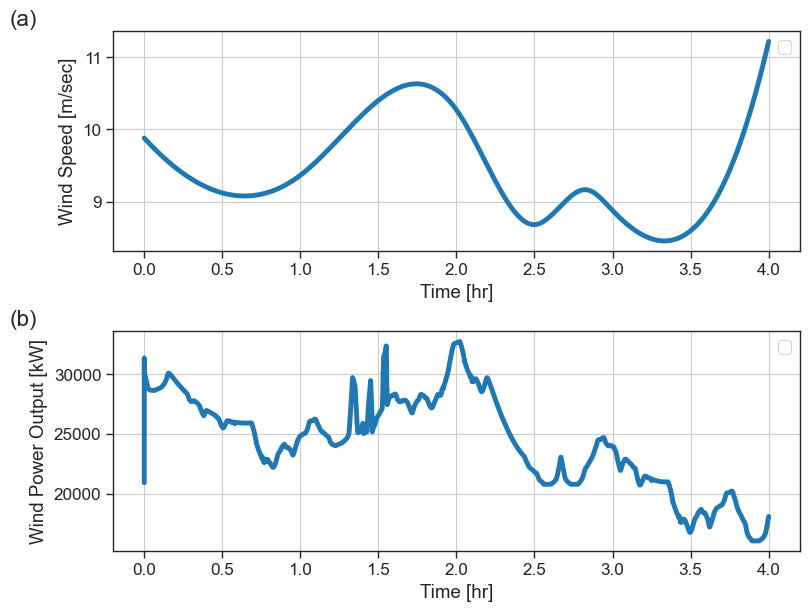}
        \caption{Wind plant outputs and resource availability during discharging.}
        \label{fig:output_WindProfile_discharging}
    \end{subfigure}

    \caption{Hybrid power plant operation under battery discharging conditions. (a) Overall plant output and control response, (b) solar subsystem behavior, and (c) wind subsystem behavior.}
    \label{fig:discharging_2x2layout}
\end{figure}

\noindent \textit{Discharging scenario with increased demand:} On the other hand, to showcase the hybrid supervisory controller's performance in discharging mode, the power demand has been increased and variations in the \textit{REG-A} signal occurs now close to $75$ to $80$ MW. For the discharging, mode we considered the following objective:
\begin{equation}\label{eq:cost_charging_01}
    \ell(P_{w_l}, P_{s_l}, P_{b_l}) = Q_r \frac{1}{2}(P_{w_l} + P_{s_l} + P_{b_l} - P_r)^2 + Q_b (P_{b_l})
\end{equation}
where the second term minimizes the contribution of power support from battery to loads such that all the available solar and wind resources can be properly utilized. Tuned gains are set at $Q_r = 10$, and $Q_b = 80$. An additional constraint on the control actions makes sure the following is satisfied:
\begin{align}
    P_{w_{d_b}} + P_{s_{d_b}} = 0
\end{align}
That is, the control commands to charge the batteries from wind and solar are set to zero.
Figure \ref{fig:output_discharging} shows that the individual subsystems can meet the desired power demand where the battery has been discharged from $20$ MW for the grid, and the solar plant is supporting at an elevated level utilizing extended availability of resources, as shown in Fig. \ref{fig:output_SolarProfile_discharging}, with the rest coming from the wind plant as shown in Fig. \ref{fig:output_WindProfile_discharging}. As the battery supports a part of the load demand, the burden on the wind and solar plants is not increased extensively, although the power demand has been increased.  In this scenario, we consider a different wind speed profile that consist of valleys and peaks to test the robustness of the feedback optimization controller. The supervisory controller intelligently manipulates the set-points of the individual sub-systems extracting the maximum possible support from the solar and wind, and at the same time, efficiently utilizes the discharging mode of the battery.

\section{Conclusions}
\label{sec:conclusion}
In this paper, we presented a supervisory controller for a hybrid plant. The supervisory control was developed to account for large uncertainties in the weather forecasting that may hinder the performance of conventional open-loop, predictive, dispatch methods for controlling hybrid plants. We use a novel control method known as feedback optimization to account for unknown weather changes by means of measurement feedback. The proposed supervisory control takes advantage of power measurements as well as wind speed and irradiance measurements to determine the maximum available power in order to coordinate the hybrid plant to meet the power demands. The proposed control was implemented in a preliminary simulation to showcase its ability to not only meet the power demands in the presence of uncertain weather, but also adjust its behavior using simple changes in the cost function to favor the use of the battery over the wind/solar farms to reduce operational costs.

The supervisory control shows promise in effectively controlling a hybrid plant. Future work will focus on updating the control architecture to address the discharging/charging cycle. Currently, the supervisory control is implemented in either a discharging/charging mode, but can be updated online if there is sufficient energy from the wind or sun to charge the battery. The switching between discharging and charging could possibly be integrated into the feedback optimization controller by formulating it as a mixed integer program. Alternatively, a planning layer could be combined with the feedback optimization control to consider discharging/charging modes over long horizons. Furthermore, the proposed supervisory control has been  integrated with the hybrid plant co-simulation engine Hercules to show the efficacy of the controller on more realistic simulations.

\section*{Author Contribution}
PNNL authors S. Mukherjee, H. Sharma, W. Cortez, and S. Glavaski contributed to the formulation, modeling, and implementation of the supervisory controller with the co-simulation engine. NREL authors G. Starke, M. Sinner, B.J. Stanislawski, Z. Tully, and P. Fleming contributed to hybrid plant configuration, resource data generation and integration with co-simulation engine. 

\section*{Declaration of Competing Interest}
There are no competing interests



\section*{Acknowledgement}

The research is funded by the U.S. Department of Energy Office of Energy Efficiency and
Renewable Energy's Wind Energy Technologies Office for the project \textit{Path to Nationwide Deployment of Fully Coupled Wind-Based Hybrid Energy Systems}. 

This work was authored in part by the by the Pacific Northwest National Laboratory, operated by Battelle Memorial Institute under contract no. DE-AC05-76RL01830 for the U.S. Department of Energy, and in part by the National Renewable Energy Laboratory for the U.S. Department of Energy (DOE) under Contract No. DE-AC36-08GO28308.  The views expressed
in the article do not necessarily represent the views of the DOE or the
U.S. Government. The publisher, by accepting the article for publication, acknowledges that the U.S. Government retains a nonexclusive, paid-up, irrevocable, worldwide license to publish or reproduce the published form of this work, or allow others to do so, for U.S. Government purposes.

\bibliography{main.bib}
\end{document}